# Revealing the nature of magnetic phases in the semi-Heusler alloy $Cu_{0.85}Ni_{0.15}MnSb$

Madhumita Halder,[1,*] K. G. Suresh,[1,#] M. D. Mukadam,[2] and S. M. Yusuf[2,†]

[1]*Department of Physics, Indian Institute of Technology Bombay, Mumbai 400076, India*

[2]*Solid State Physics Division, Bhabha Atomic Research Centre, Mumbai 400085, India*

## Abstract

We report the magnetic, magnetocaloric, and magnetotransport properties of the semi-Heusler alloy $Cu_{0.85}Ni_{0.15}MnSb$, which exhibits coexistence of antiferromagnetic (AFM) and ferromagnetic (FM) phases. A broad magnetic phase transition is evident from the temperature variations of magnetization, heat capacity, and isothermal magnetic entropy change. This is due to the presence of both AFM and FM phases at low temperatures. The variation of electrical resistivity with temperature shows three distinct regions of magnetic phases. The magnetoresistance (MR) results also show the presence of AFM and FM phases at temperatures below 45 K, and a FM phase at temperature above 45K. Though there is no signature of a spin-glass state at low temperatures, various results point towards the presence of short-range magnetic correlations at low temperatures.

Email: [*]mhalder@phy.iitb.ac.in, [#]suresh@phy.iitb.ac.in, [†]smyusuf@barc.gov.in

## 1. Introduction

In recent years, Heusler and semi-Heusler alloys have been investigated both theoretically and experimentally due to their various interesting physical properties such as half metallic ferromagnetism, shape memory effect, magnetocaloric effect, *etc*. that are promising for future technological applications [1-4]. The behavior of these systems varies from itinerant to localized magnetism with a rich diversity in the magnetic properties. The semi-Heusler alloys *X*MnSb ($X$ = 3$d$) have large local moments on the Mn atoms. The magnetic properties of these compounds depend strongly on the nature of 3$d$ ($X$) elements. The exchange interaction in these systems can be an antiferromagnetic (AFM) superexchange interaction or ferromagnetic (FM) Ruderman-Kittel-Kasuya-Yosida (RKKY) type interaction depending on the 3$d$ atoms [5].

The semi-Heusler alloy NiMnSb is a ferromagnet with a Curie temperature ($T_C$) of 750 K, and crystallizes in the $C1_b$ structure with four interpenetrating fcc sublattices [6]. In this compound, the exchange interaction is mainly of RKKY type [7]. On the other hand, though the semi-Heusler alloy CuMnSb has the same crystal structure as that of NiMnSb, it is antiferromagnetic with a Néel temperature ($T_N$) of 55 K and the exchange interaction is of superexchange type [8]. It has been observed that there is a transition from AFM to FM phase in $Cu_{1-x}Ni_xMnSb$ with increase in Ni concentration [9-11]. This transition occurs due to the change in electron concentration (i.e. difference in Cu and Ni valences), which modifies the density of states at the Fermi level [12, 13]. This affects the exchange interaction between Mn-Mn spins in the Mn sublattice, resulting in the AFM to FM transition. It is also observed that there is a region of magnetic phase coexistence in $Cu_{1-x}Ni_xMnSb$ series [9]. In an earlier report [9], we observed that for $x < 0.05$, $Cu_{1-x}Ni_xMnSb$ is mainly in the AFM state. In the region $0.05 \leq x \leq 0.2$, there is coexistence of AFM and FM phases. For $x > 0.2$, the system is in the ferromagnetic state. For $x = 0.15$, we observed the coexistence of long-ranged AFM and FM phases. Therefore, this concentration presents an interesting scenario, which is expected to result in interesting magnetic and related properties such as magnetocaloric effect and magnetoresistance. In the present article, we report the results of these studies on this alloy. It is also expected that this composition might exhibit spin-glass behavior in the coexistence region. Since there is coexistence of both AFM and FM phases, a short-range magnetic correlation is also expected to exist in this coexistence region. A detailed neutron diffraction study has been carried out to explore this possibility.

## 2. Experimental procedure

The polycrystalline sample of $Cu_{0.85}Ni_{0.15}MnSb$ was prepared by the arc-melting method as described in our previous report [9]. The dc magnetization measurements were carried out on the sample using a Physical Property Measurement System (PPMS, Quantum Design) as a function of temperature and magnetic field. The zero-field-cooled (ZFC) and field-cooled (FC) magnetization measurements were carried out over the temperature range of 5-300 K under 200 Oe field. Magnetization ($M$) as a function of magnetic field ($H$) was measured at 5 and 30 K over a field variation of ± 90 kOe. The magnetization isotherms were recorded at various temperatures with an interval of 5 K up to a maximum applied field of 50 kOe. The ac susceptibility $\chi'_{ac}$ and heat capacity measurements were carried out using the PPMS. $\chi'_{ac}$ measurements were carried out at frequencies from 7 Hz to 993 Hz with an applied ac field amplitude of 1 Oe. Electrical resistivity and magnetoresistance were measured using the standard four probe method. The temperature dependent neutron diffraction experiments were performed down to 1.5 K on the neutron powder diffractometer DMC (λ = 2.4585 Å) at the Paul Scherrer Institute (PSI), Switzerland.

## 3. Results and discussions

Figure 1 shows ZFC and FC *M* vs temperature (*T*) curves under an applied field of 200 Oe, for the $Cu_{0.85}Ni_{0.15}MnSb$ sample. A broad antiferromagnetic peak at 45 K, with a clear bifurcation in the ZFC and FC magnetization curves below this temperature has been observed. This bifurcation is due to the presence of a competing AFM and FM interactions. The inset of Fig. 1 shows the *M* vs *H* curves at 5 and 30 K over a field range of ± 90 kOe (i.e. all four quadrants). There is no field induced transition even up to a field of 90 kOe. This indicates that the AFM phase does not undergo any metamagnetic like transition to a ferromagnetic phase even at high fields. The saturation magnetization at 5 K and 90 kOe is 2.5$\mu_B$/f.u. The moment is localized on Mn atoms both in the case of NiMnSb [14] (3.85 $\mu_B$/Mn atom) and CuMnSb [15] (3.9 $\mu_B$/Mn atom). The value of saturation magnetization in the present case is less as the volume fraction of FM phase only contributes to the magnetization. A negligible hysteresis has been observed at 5 K. Following the well developed procedure [16, 17], the magnetocaloric effect in terms of isothermal magnetic entropy change ($\Delta S_M$) was estimated from the magnetization isotherms using the equation

$$\Delta S_M(H,T) = \int_0^H \left( \frac{\partial M(H,T)}{\partial T} \right)_H dH \qquad (1).$$

For magnetization data collected at discrete temperature intervals and fields, $\Delta S_M$ (*H, T*) can be approximated as

$$\Delta S_M(H,T) = \sum_i \frac{M_{i+1}(T_{i+1},H) - M_i(T_i,H)}{T_{i+1} - T_i} \Delta H \qquad (2).$$

Figure 2 shows the variation of -$\Delta S_M$ with temperature. It may be noted that the entropy change is negative in this case. For field variations of 10, 30, and 50 kOe, the maximum values of -$\Delta S_M$ are found to be 0.4, 1.0, and 1.5 J $kg^{-1}$ $K^{-1}$, respectively. The value of -$\Delta S_M$ almost remains constant over a wide temperature range and decreases gradually as the temperature shifts away from the magnetic transition temperature (Fig. 2). This is because the magnetic transition spreads over a broad temperature range as observed in the *M* vs *T* curve (Fig. 1). Though the value of -$\Delta S_M$ is not large, it has a wide operating temperature range with negligible hysteresis which is important for a practical magnetic refrigerator. We would like to mention here that one would expect $\Delta S_M$ to be positive below the AFM transition temperature. But since the sample contains fixed AFM and FM phases with the latter as the dominant component with application of field, results in negative $\Delta S_M$. The nature of magnetic transition (i.e., paramagnetic to FM transition) for the present $Cu_{0.85}Ni_{0.15}MnSb$ sample has been studied by the Arrott plot ($M^2$ vs $H/M$ curves) method. According to the Banerjee criterion [18], a positive slope in the $M^2$ vs $H/M$ curves indicates a second-order transition, while a negative slope indicates a first-order transition. Only a positive slope of the $M^2$ vs $H/M$ curves near the magnetic transition has been observed (shown in Fig. 3) in the present case, which indicates that the magnetic transition (i.e paramagnetic to FM transition) is of second order in nature. In order to investigate the phase coexistence region in detail, we have carried out the heat capacity measurements. Figure 4 shows the temperature dependence of heat capacity (*C*) in zero field and in 30 kOe. A small kink is observed at around 43 K corresponds to the AFM transition. Absence of any sharp peak at AFM transition indicates that this transition is weak. There is no signature of any abrupt change in the value of *C* with temperature above 43 K, indicating that the change in the magnetic entropy during the FM transition is quite small. This is also evident from *M* vs *T* curve and the small change in $\Delta S_M$ in the FM region. The adiabatic temperature change ($\Delta T_{ad}$), calculated using the equation (3), is found to be 0.5 K for a field of 30 kOe

$$\Delta T_{ad} = -\int_0^H \frac{T}{C_H}\left(\frac{\partial M}{\partial T}\right)_H dH \qquad (3).$$

The broad transition could be realized on the basis of the results of our earlier neutron scattering study [9] which showed the coexistence of AFM and FM phases at low temperatures. Above 45 K, the AFM phase is transformed into paramagnetic phase, but the FM phase still persists. This leads to the broad transition which is observed above 45 K. With further increase in temperature the system finally goes to paramagnetic state [9]. In order to probe the existence of any spin glass state, we have carried out the ac susceptibility measurements. Figure 5 depicts the temperature dependence of the real part of the ac susceptibility ($\chi'_{ac}$) at various frequencies. A peak appears at around 50 K in the $\chi'_{ac}$ vs $T$ plot which denotes the AFM transition temperature. No shift in the peak position was observed with change in ac excitation frequency, indicating that there is no glassy behavior present in the sample. However, a sluggish nature of magnetic transition is evident (inset of Fig. 5). From the heat capacity and the ac susceptibility measurements, we can conclude that both AFM and FM transitions are quite weak.

To further investigate the phase coexistence region, we have carried out electrical resistivity and magnetoresistance measurements. The variation of resistivity ($\rho$) with $T$ measured under various fields (including zero field) is depicted in Fig. 6. The magnitude of resistivity shows that the sample is in the metallic state. There is a clear change in the slope of $\rho$ vs $T$ curves at around 45 K and 160 K. These two temperatures corresponds to the AFM and FM transition temperatures, respectively. As mentioned earlier, below 45 K, the $Cu_{0.85}Ni_{0.15}MnSb$ sample is in the mixed AFM and FM phases while in the temperature region 45 K < $T$ < 160 K, the sample is in FM phase. Above 160 K the sample goes to its paramagnetic state. The resistivity of a ferromagnetic material can be described by the following relation [19]

$$\rho(T) = \rho_0 + \rho_{\mathrm{ph}}(T) + \rho_{\mathrm{mag}}(T) \qquad (4).$$

Here $\rho_0$ is the temperature independent residual resistivity that arises from the scattering of conduction electrons by lattice defects, domain walls, *etc*. $\rho_{ph}$ and $\rho_{mag}$ are the contributions from the electron-phonon and electron-magnon scatterings, respectively. These are temperature dependent and the dominance of $\rho_{ph}$ or $\rho_{mag}$ depends on the temperature region under consideration. In the magnetically ordered state, the temperature dependent part of the resistivity is due to electron-magnon, electron-phonon, and electron-electron scattering. The temperature variation of electron-electron and electron-magnon scattering shows a $T^2$ dependency [20, 21]. In the FM region i.e. 45 K < $T$ < 160 K, we have fitted the data (inset of Fig. 6) to the following expression

$$\rho(T) = \rho_0 + \mathrm{A}T + \mathrm{B}T^2 \qquad (5).$$

Here, the terms A$T$ and B$T^2$ are due to electron-phonon and electron-magnon scattering, respectively. The fitted values of A and B are $8.5\times10^2$ n Ω cm $K^{-1}$ and 2.0 n Ω cm $K^{-2}$, respectively. In the temperature region below 45 K, there are mixed AFM and FM phases and so we have fitted the data in this region to $T^2$ and $T^n$ i.e., with $\rho(T) = \rho_0 + \mathrm{A}T^2 + \mathrm{B}T^n$, the term $T^n$ is used for the AFM phase [22]. Since at low temperatures, the contribution of electron-phonon scattering is less as compared to other scatterings [23], we have neglected the A$T$ term. The fitted curve in the temperature range 7 - 42 K is shown in the inset of figure 6. The value of $n$ is found to be 1.7 which is close to that found for the parent CuMnSb [22]. The fact that $n$ is not equal to 2 suggests the presence of an AFM phase.

The MR which denotes the percentage change of resistivity due to applied magnetic field is defined as $[\rho(H) - \rho(0)]/\rho(0) \times 100$. Figure 7 shows the MR estimated at different temperatures for applied fields up to 90 kOe. MR is negative at high temperatures and attains a maximum value of 6 % for a field of 90 kOe at 100 K. In the low temperature region, i.e., below 45 K, the MR is positive in low fields and becomes negative in high fields. In the paramagnetic region, there is a linear increase in MR with field. In the ferromagnetic region, the change in MR with field is found to be more pronounced. In the low temperature region, where AFM and FM phases coexist, the MR is positive at low fields and becomes negative at high fields. A similar behavior was reported for $Fe_{3-x}Mn_xSi$ alloys with competing ferro and antiferromagnetic interactions [24]. In the case of ferromagnetic and paramagnetic states with localized magnetic moments, the magnetic field suppresses the spin fluctuations, which leads to a negative magnetoresistance [25]. Yamada *et al*[26] have theoretically shown that a positive magnetoresistance can arise in an antiferromagnetic system with localized magnetic moments. They showed that in a polycrystalline AFM material, with increase in applied field, MR changes from positive to zero, and then becomes negative. In the case of present $Cu_{0.85}Ni_{0.15}MnSb$ sample, only Mn atoms carry the moment [9] and it is a localized magnetic moment system. Hence the above argument can be used to describe the temperature and the field dependencies of MR of the system. At temperature below 45 K, since the sample has mixed FM and AFM phases, the AFM phase causes the positive MR at low fields and negative MR at high fields. On the other hand, the FM phase leads to a negative MR, thus resulting in a small positive MR in the low field and low temperature regime. Since there is no metamagnetic-like transition of the AFM phase with externally applied fields, the negative value of MR at high fields can be attributed to the suppression of the AFM spin fluctuations.

In order to throw more light into the magnetic state of $Cu_{0.85}Ni_{0.15}MnSb$, we have performed a neutron diffraction study at different temperatures. In our earlier study, we reported the coexistence of both AFM and FM phases in $Cu_{1-x}Ni_xMnSb$ system [9]. From the diffraction pattern for $Cu_{0.85}Ni_{0.15}MnSb$ sample, on careful analysis, we also observe a broad hump (shown in Fig. 8) at low temperature and in the low $Q$ region ($Q = 4\pi \sin\theta/\lambda$). This indicates a possible existence of short-range magnetic ordering present in the sample. This hump is prominent at low temperature 1.5 K and 50 K. As the temperature is increased to 200 K, this broad hump almost vanishes. It is to be noted that this hump occurs at a $Q$ position ($Q$ = 0.64 Å$^{-1}$) which is not allowed under the $F\bar{4}3m$ space group. This indicates the presence of some short-range magnetic correlation or clustering among the FM and AFM phases. Thus, apart from the coexistence of long-ranged AFM and FM phases, there is also the existence of short-range magnetic correlation. This short-range magnetic correlation is more prominent in the low temperature region, where there is coexistence of AFM and FM phases. As the temperature increases, the AFM phase disappears and only the FM phase persists so that the hump is less prominent. At high temperature i.e., in the paramagnetic region, the broad hump disappears.

## 4. Summary and conclusion

In this paper, we have investigated the magnetic and related properties of $Cu_{0.85}Ni_{0.15}MnSb$ alloy in detail. The sample shows a broad second order magnetic transition, reflected in the magnetic entropy change and the heat capacity data, which is due to the presence of both AFM and FM phases. No spin glass behavior is observed for the sample; but there is an evidence of some short-range magnetic correlation, which is more prominent at low temperatures ($T \leq 50$ K). The variation of electrical resistivity with temperature shows three distinct regions of magnetic phases, viz. (i) $T \leq 45$ K with existence of both AFM and FM phases, (ii) $45 \leq T \leq 160$ K region with mixed FM and paramagnetic states, and (iii) $T \geq 160$K region in the paramagnetic state. The magnetoresistance study also confirms the presence of AFM and FM phases at low temperatures below 45 K, and a FM phase above 45K. Above 160 K the sample is in the paramagnetic phase. The results of the electrical resistivity and magnetoresistance are in agreement with magnetization and neutron diffraction results. The present results are quite useful in understanding the behavior of other systems with two competing magnetically ordered phases and can possibly help in the tuning the ferromagnetic component in other Heusler alloys which are otherwise AFM in nature. This can be of great use in many applications.

**Acknowledgment**

The authors would like to acknowledge the help provided by L. Keller, Paul Scherrer Institute, Villigen, Switzerland for the neutron diffraction measurements. M. H. thanks CSIR, India for the fellowship.

Refrences

**List of Figures**

Fig. 1: (Color online) Temperature dependence of FC and ZFC magnetization (*M*) for $Cu_{0.85}Ni_{0.15}MnSb$ at 200 Oe. Inset shows *M* vs *H* curve over all the four quadrants at 5 and 30 K.

Fig. 2: (Color online) Magnetic entropy change $-\Delta S_M$ vs *T* for $Cu_{0.85}Ni_{0.15}MnSb$.

Fig. 3: $M^2$ vs *H*/*M* isotherms at different temperatures for $Cu_{0.85}Ni_{0.15}MnSb$.

Fig. 4: (Color online) Heat capacity for $Cu_{0.85}Ni_{0.15}MnSb$ as a function of temperature in zero field and 30 kOe. Arrow marks the AFM transition.

Fig. 5: (Color online) Temperature dependence of the real part of ac susceptibility at various frequencies with an ac field of 1 Oe for $Cu_{0.85}Ni_{0.15}MnSb$. The inset shows the enlarge region near the transition temperature.

Fig. 6: (Color online) Temperature dependence of electrical resistivity at various fields for $Cu_{0.85}Ni_{0.15}MnSb$. The inset shows the temperature variation of electrical resistivity in zero field fitted to $\rho(T) = \rho_0 + AT^2 + BT^n$ and $\rho(T) = \rho_0 + AT + BT^2$ in the temperature range 7 - 42 K and 48 - 155 K, respectively.

Fig. 7: (Color online) Field dependence of magnetoresistance (MR) at various temperatures for $Cu_{0.85}Ni_{0.15}MnSb$.

Fig. 8: (Color online) Low Q neutron diffraction pattern at different temperature for $Cu_{0.85}Ni_{0.15}MnSb$.

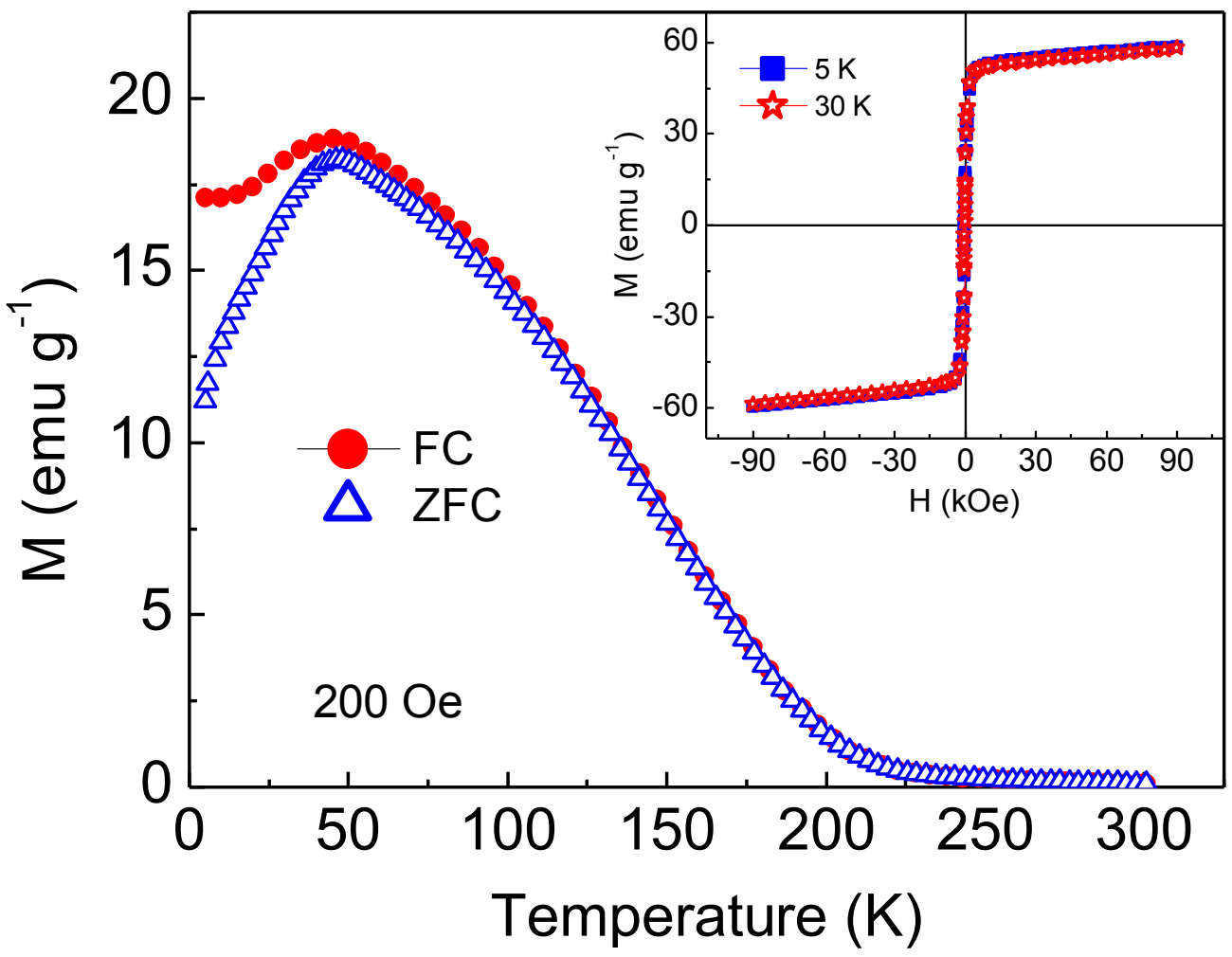

Fig. 1

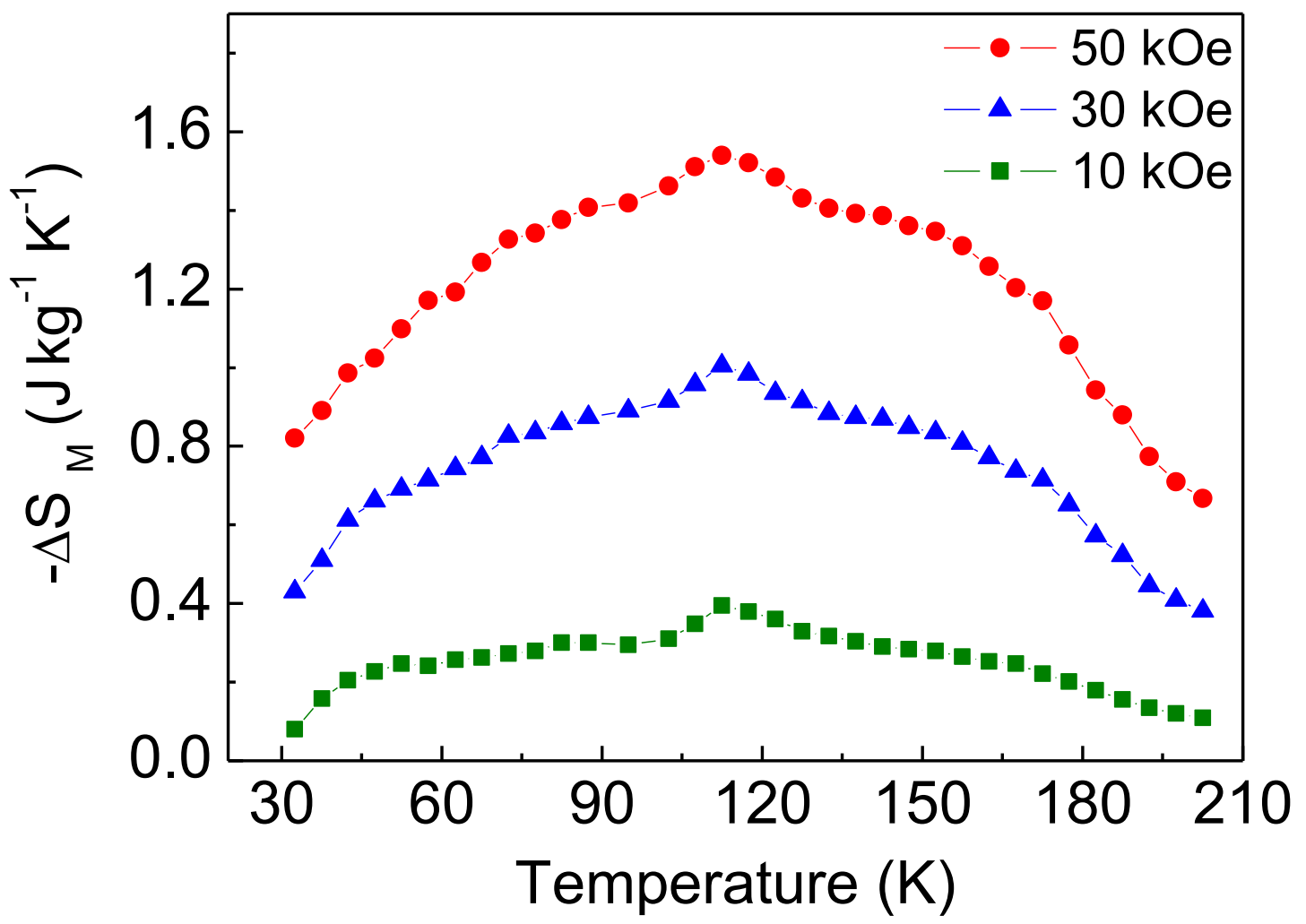

Fig. 2

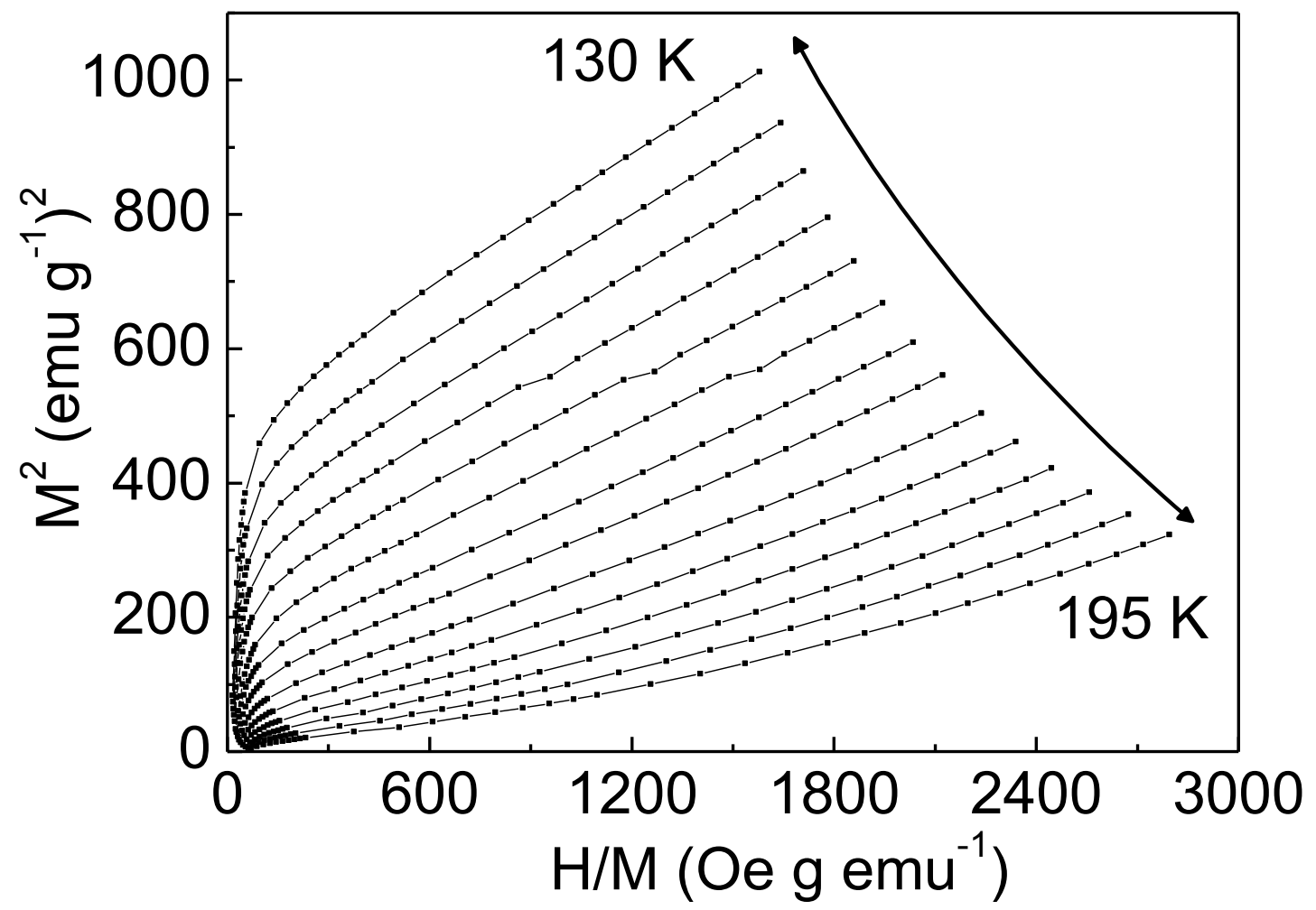

Fig. 3

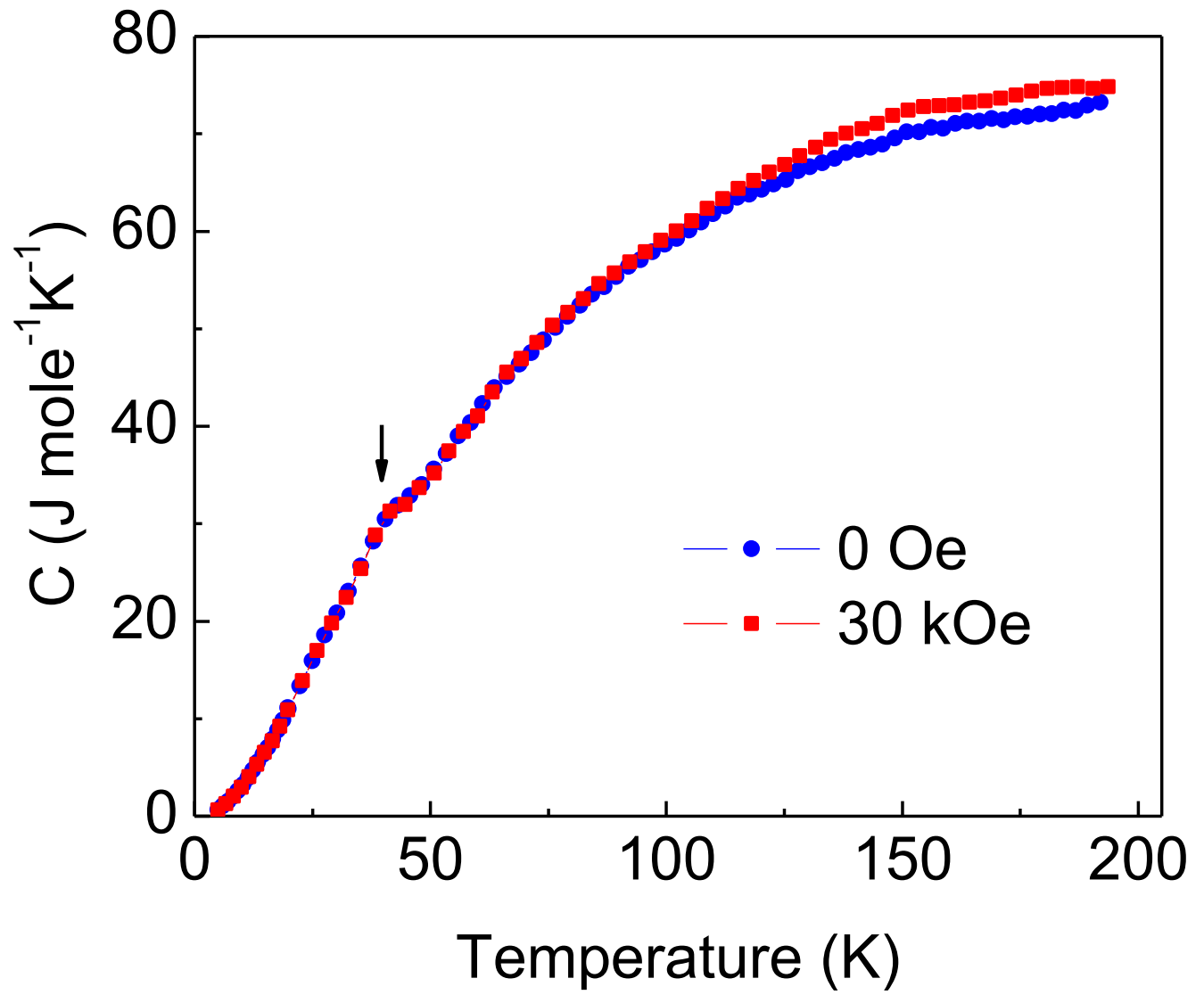

Fig. 4

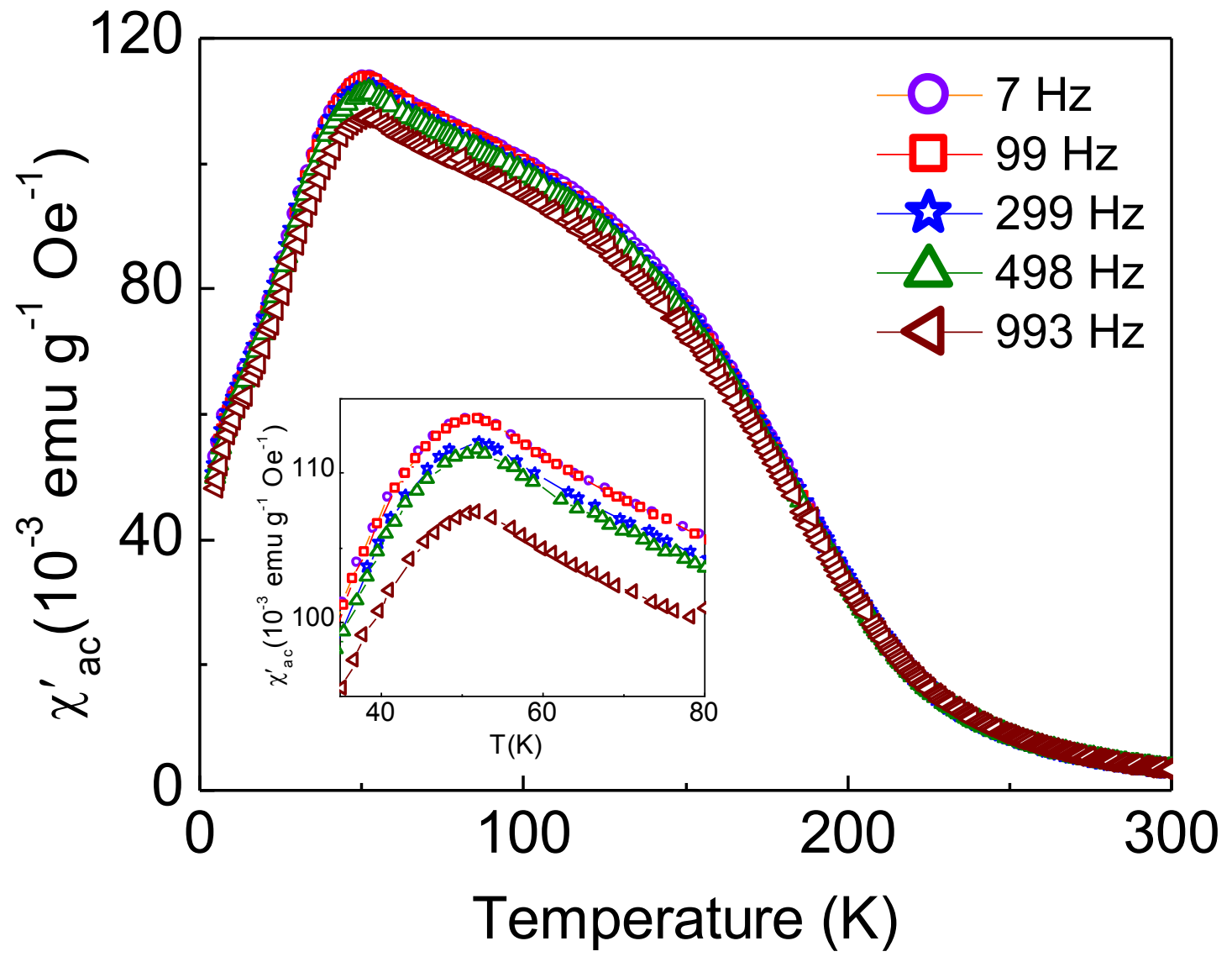

Fig. 5

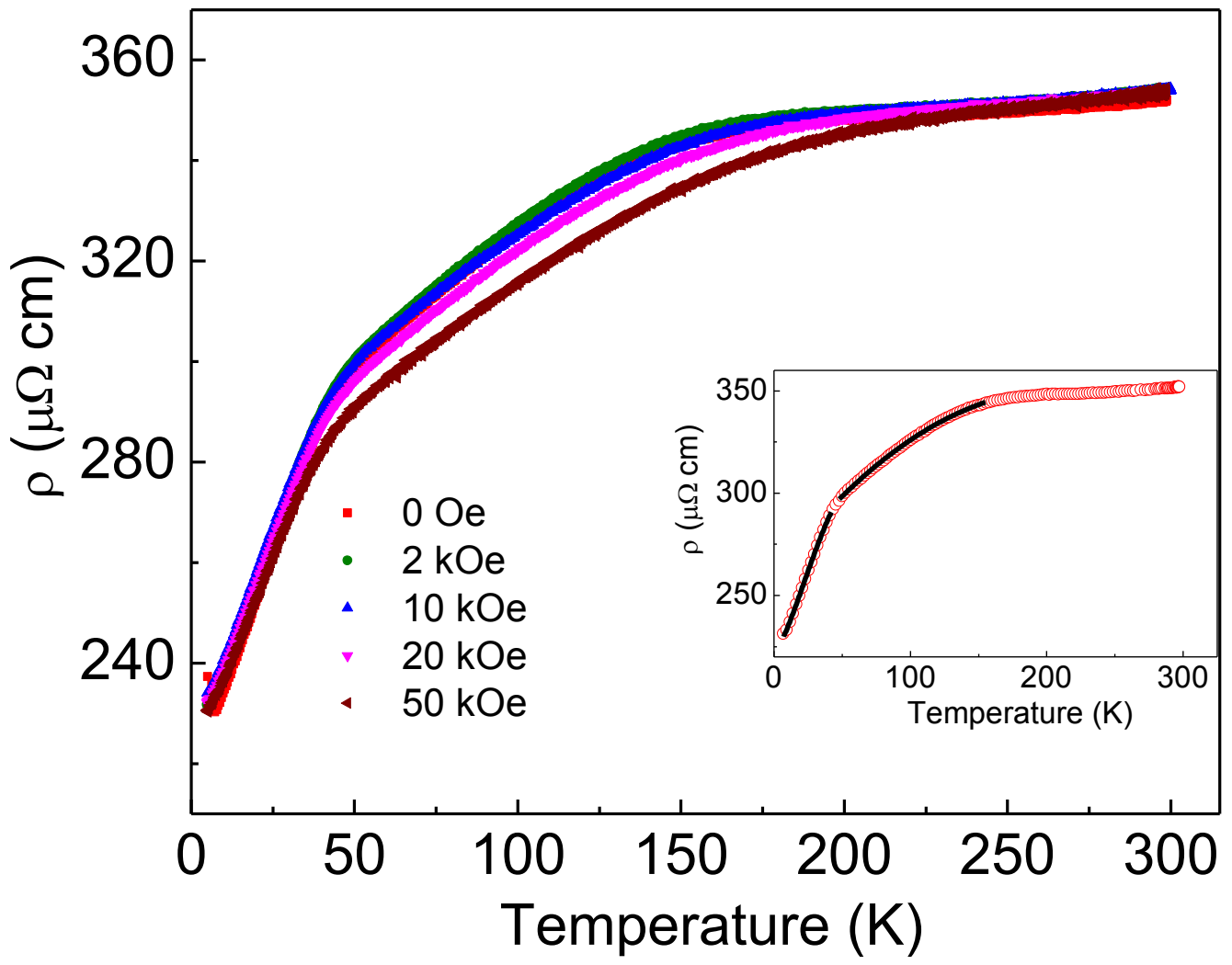

Fig. 6

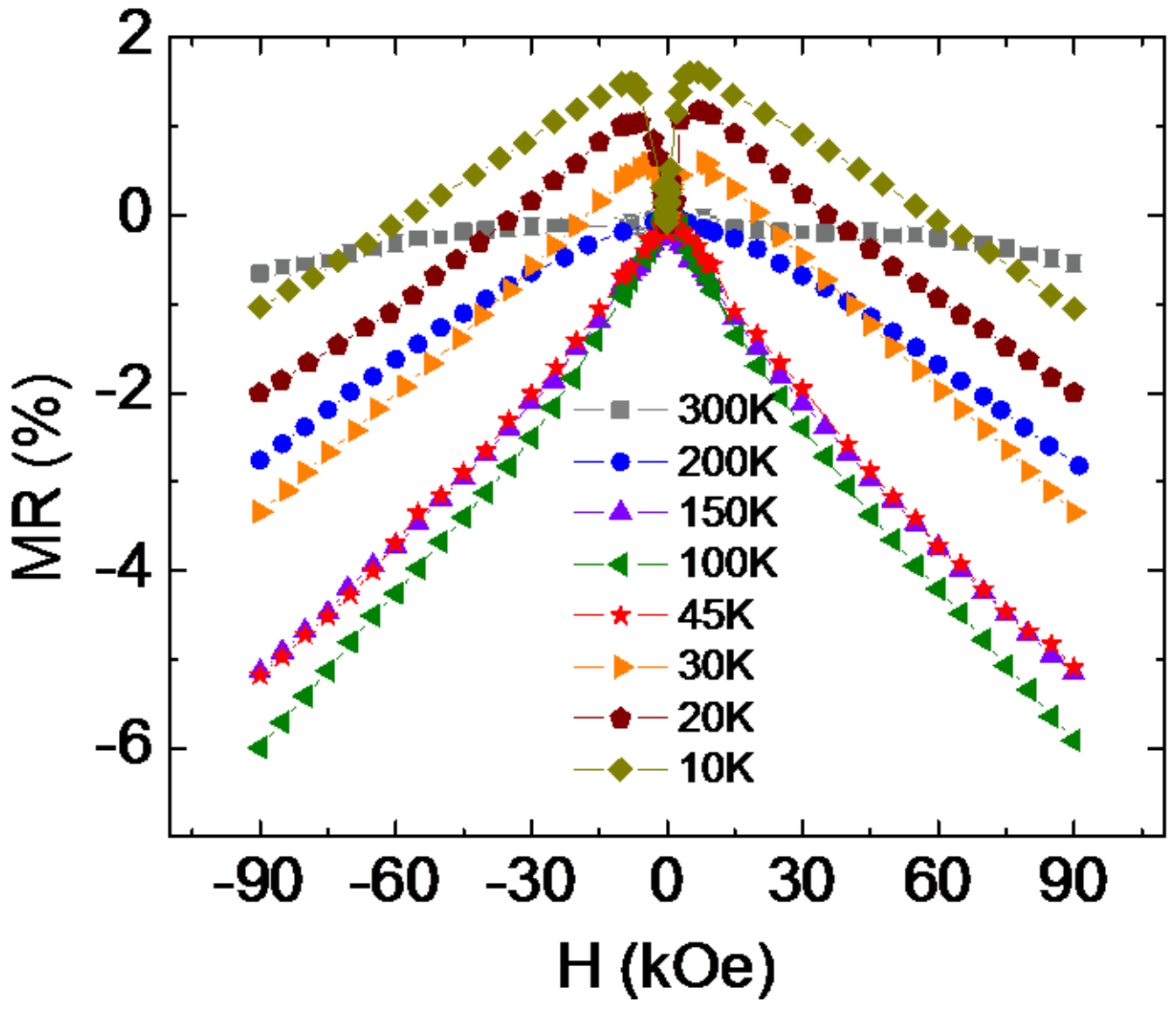

Fig. 7

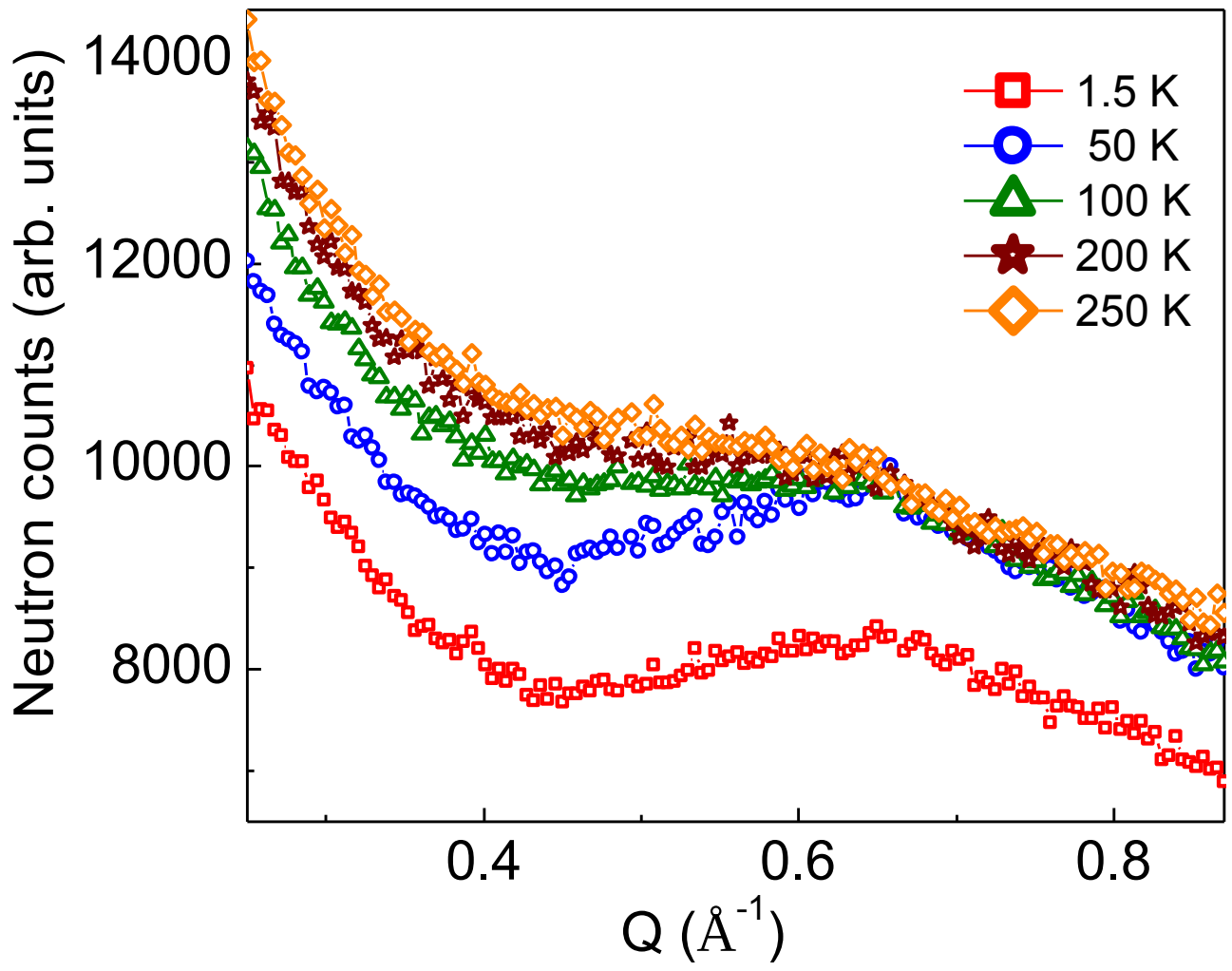

Fig. 8